\documentstyle[11pt]{article}

\title{\bf The photon-photon interaction at low $x$ in the theory of
reggeized gluons with a running coupling and
 $N_{c}\rightarrow\infty$}
\author{M.Braun \\ Department of high-energy
physics, \\ University of S. Petersburg, 198904 S. Petersburg
, Russia}
\def\beq{\begin{equation}}
\def\eeq{\end{equation}}
\def\noi{\noindent}

\begin{document}
\maketitle
\medskip
\noi{\bf Abstract.}

The forward elastic amplitude for scattering of real and weakly virtual
photons is studied in the framework of the theory of reggeized gluons,
with large $N_{c}$ and a running coupling constant introduced in the
manner which preserves the bootstrap condition. Transition from a single
to multiple pomeron exchanges is observed as $x$ gets smaller. For very
low $x$ the amplitude aquires an eikonal form. The photon structure
function reveals a strong violation of scaling: it grows as $Q^{2}$.
As a function of $x$ it behaves as $(\ln(1/x)\ln\ln(1/x))^{2}$.
Correspondingly the
cross-section for physical photons grows with energy as
$(\ln s\ln\ln s)^{2}$. Hadronic structure functions and cross-sections
are also briefly discussed.\vspace{3 cm}

{\Large\bf SPbU-IP-1995/7}
\newpage
\section{Introduction}

As has been shown in [ 1 ], in the framework of the "hard pomeron"
approach of BFKL  with a fixed small coupling constant [ 2 ], the
physical
picture of the soft strong interaction drastically simplifies in the
limit of infinite number of colours $N_{c}\rightarrow\infty$ and
reduces to an exchange of arbitrary number of non-interacting BFKL
pomerons. Taking photons as projectile and target (there are no hadrons
in the BFKL theory with a small coupling constant) and summing over all
multipomeron exchanges one arrives at an eikonal amplitude, very similar
to the old supercritical multipomeron exchange ("Froissaron") model [ 3
], except that the pomeron is composite (and therefore nonlocal)
and does not involve any scale. These
circumstances together with the properties of the photon colour density
lead to a behaviour of the total cross-section very different from the
Froissaron model. The total cross-section continues to rise as a power
of energy, although somewhat smaller than with a single pomeron
exchange. Also a strong violation of scaling is observed in the
structure function: it rises as $Q$.

To relate these results to the physical world and realistic QCD one has
at least to introduce a running coupling and thus a scale. There is
some hope that the  strong violation of the Froissart unitarity
bound by a unitary amplitude observed in [ 1 ] may be explained by the
absence of scale and an effectively zero gluon mass. The results of [ 1
] are heavily based on the so-called bootstrap condition [ 4 ], which
allows to eliminate all nontrivial gluonic configurations in the limit
$N_{c}\rightarrow\infty$. So the only reasonable method to introduce the
running coupling seems to be the one which we proposed some time ago,
based on the bootstrap equation and thus preserving it [ 5 ]. In the
theory with the running coupling introduced according to [ 5 ]
the only nontrivial configurations which remain in the limit
$N_{c}\rightarrow\infty$ are non-interacting pomerons , as in the case
of a fixed coupling constant discussed in [ 1 ]. The photon-photon
amplitude which results after summing over all multipomeron exchanges
is then  essentially the same as obtained in [ 1 ], except that now the
pomeron has to be calculated with a running coupling, as in [ 5 ].
It now posseses a nontrivial dimensionful slope $\alpha'$ and also
may be a pole in the complex angular momentum plane rather than a cut.

The purpose of the present note is to study the behaviour of the
photon-photon cross-section and structure function obtained with this
pomeron at high energies and low $x$ and to compare it to the scaling
result of [ 1 ]. The main conclusion is that
 as a function of energy $\sqrt{s}$ the cross-section  now behaves as
\beq
\sigma^{tot}\sim(\ln s)^{2}(\ln\ln s)^{2}
\eeq
Thus the rise with energy has become much slowlier. However the
Froissart bound still remains violated (although at extraordinary high
energies). We expect this violation to be cured if higher orders in
electromagnetic coupling are taken into account.

\section{The photon-photon amplitude with a running coupling}

We consider elastic scattering of two transverse photons with momenta $q$
(projectile) and $p$ (target) one of which or both may be virtual
($q^{2}=-Q^{2},\ p^{2}=-P^{2}$). With the number of colours
$N_{c}\rightarrow\infty$ and a fixed small coupling constant $g$ the
forward amplitude obtained after summing over all pomeron exchanges has
an eikonal form [ 1 ]
\[
A(q,p)=4i\nu e^{4}N_{c}^{2}
\]\beq
\int\,d^{2}R \int d^{2}r \int d^{2}r'(2\pi)^{-6}
\rho_{q}(r)\rho_{p}(r')
(1-\exp (-z(\nu,R,r,r')))
\eeq
Here $\nu=pq\rightarrow\infty$, $\rho_{q,p}(r)$ describe the
dipole colour density of the projectile ($q$) and target ($p$) photons
as a function of the transverse radius of the $q\bar q$ pair into which
the photons decay. For the projectile
\beq
\rho_{q}(r)=
\sum_{f=1}^{N_{f}}Z_{f}^{2}\int_{0}^{1}d\alpha
(m_{f}^{2}{\rm K}_{0}^{2}(\epsilon_{f} r)
+(\alpha^{2}+(1-\alpha)^{2})\epsilon_{f}^{2}{\rm K}_{1}^{2}
(\epsilon_{f} r))
\eeq
where $\epsilon_{f}^{2}=Q^{2}\alpha (1-\alpha)+m_{f}^{2}$ and $m_{f}$
and $Z_{f}$ are the mass and charge of the quark of flavour $f$.  The
target density has the same form with $Q\rightarrow P$. At small $r$
these densities behave as $1/r^{2}$:
\beq
\rho_{q}(r)\sim (2/3)r^{-2}\sum_{f}Z_{f}^{2},\ \ Qr\rightarrow 0
\eeq
The eikonal function $z(\nu,R,r,r')$ represents the contribution of a
single pomeron exchange
\[z(\nu,R,r,r')=(g^{4}/16)\int\,(d^{2}qd^{2}q_{1}d^{2}q'_{1}/(2\pi)^{6})
f(\nu,q,q_{1},q'_{1})\exp(iqR)\]\beq (\exp iq_{1}r -1)
 (\exp iq_{2}r -1)(\exp (-iq'_{1}r') -1)
(\exp (-iq'_{2}r') -1)
\eeq
Here $q=q_{1}+q_{2}=q'_{1}+q'_{2}$
 and $g$ is the fixed QCD coupling. The
function $f(\nu,q,q_{1},q'_{1})$ is the solution of the BFKL equation
for the inhomogeneous term
$(2\pi)^{2}q_{1}^{-2}q_{2}^{-2}\delta^{2}(q_{1}-q'_{1})$

Our task is to generalize this expression to the case of a running
coupling constant. As mentioned in the Introduction, we introduce it
in the way compatible with the bootstrap condition. This means that we
change both the reggeized gluon trajectory $\omega$ and its interaction
$V$ in a specific manner compatible with the bootstrap equation. In the
BFKL theory both $\omega$ and $V$ can be expressed through combinations
$q^{2}/g^{2}$ where $q$ is some transverse momentum and $g$ the fixed
coupling constant. To introduce the running coupling, according to [ 5 ]
one has to substitute for each $q$
\beq
q^{2}/g^{2}\rightarrow\eta(q)
\eeq
where $\eta(q)$ is a certain function with the asymptotical behaviour
\beq
\eta(q)\simeq (b/8\pi)q^{2}\ln (q^{2}/\Lambda^{2})
\eeq
Here $b=(11/3)N_{c}-(2/3)N_{f}$, $N_{c,f}$ are the numbers of colours
and flavours and $\Lambda$ is the standard QCD parameter. The low
momentum behaviour of $\eta(q)$ remains undetermined and reflects the
uncertainty associated with the confinement region. One can show that
this recipe correctly reproduces the asymptotical behaviour of the
gluonic distribution at high momenta (in the double logarithmic
approximation).

With (6) the amputated wave function $\psi$ (without external legs)
satisfies a BFKL-like equation
\beq
(j-1-\omega_{1}-\omega_{2})\psi (q_{1},q_{2})=V\psi (q_{1},q_{2})
\eeq
where $j$ is the complex angular momentum,
$\omega_{1}\equiv\omega (q_{1})$ etc,
\beq
\omega=-(N_{c}/2)\eta\int\, d^{2}q_{1}(2\pi)^{-2}
\eta_{1}^{-2}\eta_{2}^{-2}
\eeq
the kernel of $V$ has the form
\beq
V(q_{1},q_{2},q'_{1},q'_{2})=-N_{c}((\eta_{1}/\eta'_{1}+\eta_{2}
/\eta'_{2})/\eta(q_{1}-q'_{1})-\eta/(\eta'_{1}\eta'_{2}))
\eeq
and $q=q_{1}+q_{2}=q'_{1}+q'_{2}$.

Turning to Eq. (5) we have first to note that the Green function $f$
involves nonamputated wave functions $\phi$, which in the fixed coupling
case are related to $\psi$ by
\beq
\phi (q_{1},q_{2})=q_{1}^{-2}q_{2}^{-2}\psi (q_{1},q_{2})
\eeq
In the amplitude these gluonic distribution functions are attached to the
external source, that is, to  the projectile  dipole colour densitiy.
The strength of this coupling is represented by the factor $g^{2}$ in
Eq. (5). The second $g^{2}$ refers to the target. The problem is to
introduce the running coupling at this point, where we have no bootstrap
relation as a guiding principle.

To do that we recall that, as shown in [ 1 ], in the photon density (3)
all intermediate quark and antiquark states appear on the mass shell
except for the initial and final ones. Let us first forget about these
exceptional states. Then each coupling constant in the $q\bar q$ loop
depends on only one momentum, that of the attached gluon $q$ and it is
evident that to introduce the running coupling one simply has to use
$g(q)$ instead of the fixed $g$. According to (6)
\beq
g^{2}(q)=q^{2}/\eta(q)
\eeq
Comparing this to (11) we find that the introduction of the running
coupling to the external source can be realized by the substitution
\beq
g^{2}\phi (q_{1},q_{2})\rightarrow\frac{\psi (q_{1},q_{2})}
{\sqrt{q_{1}^{2}q_{2}^{2}\eta_{1}\eta_{2}}}\equiv\phi (q_{1},q_{2})
\eeq

Normalizing the solutions $\psi_{\kappa}$ of  Eq. (8) according to
\beq
\int (d^{2}q_{1}/(2\pi)^{2})\psi^{\ast}_{\kappa'}(q_{1},q_{2})
\psi_{\kappa}(q_{1},q_{2})\eta^{-1}(q_{1})\eta^{-1}(q_{2})=\delta
(\kappa'- \kappa)
\eeq
we find for the running coupling case
\beq
g^{4}f(\nu,q,q_{1},q'_{1})=\int d\kappa \nu^{\alpha(\kappa)-1}
\phi_{\kappa}(q_{1},q_{2})\phi^{\ast}_{\kappa}(q'_{1},q'_{2})
\eeq
This formula together with (5) introduces the running coupling into the
eikonal factor in Eq. (2).

However we have still to account for the two mentioned exceptional quark
states in both projectile and target, which are virtual. The coupling
constant associated with them depends not only on the attached gluon
momentum $q$ but also on the quark momentum $k$. This latter dependence
has a definite influence of the high energy behaviour of the
cross-sections and cannot be neglected. It is essential only when the
quark momentum is much larger than the gluon's one when, instead of
 $1/\ln q^{2}$ taken into account in (12), $1/\ln k^{2}$ should appear.
We are not able to take this effect in a rigorous manner, since, strictly
speaking, it destroys the factorizability of multigluon exchanges making
some gluons exceptional. Very approximately we shall add to each virtual
quark propagator an extra factor $1/\sqrt{\ln k^{2}}$ at high
$k^{2}$.
Together with (6) this recipe evidently consists in approximating the
 coupling
constant of the virtual quark $g(k,q)$ as  $g(q)/\sqrt{\ln
k^{2}}$, which correctly
describes its behaviour at $k^{2}>>q^{2}$. We shall
see that these values of $k$ play the dominant role in the high-energy
behaviour of the amplitude. This change in the quark propagator results
 in
modyfying the McDonald functions entering (3)
\beq
K_{0,1}(z)\rightarrow\chi_{0,1}(z)
\eeq
where now
\beq
2\pi\chi(z)=\int\frac{ d^{2}k\exp(ikz)}{(k^{2}+1)\sqrt{\ln c(k^{2}+1)}}
\eeq
with $c>1$ and $\chi_{1}(z)=-\chi'_{0}(z)$. The dependence on $c$
reflects our ignorance of the confinement region, as with the function
$\eta$. The inclusion of an extra logarithmic factor in (17) makes the
small $r$ behaviour of the colour density $\rho (r)$ somewhat less
singular (cf (4)). Now
\beq
\rho_{q}(r)\sim (2/3)(r^{2}\ln(1/Q^{2}r^{2}))^{-1}\sum_{f}Z_{f}^{2},\ \
Qr\rightarrow 0 \eeq
However its integral over $r$ still diverges at small $r$.

\section{Photonic cross-sections at high energies}

To stay within the perturbative regime we have to choose all the momenta
in (5) much greater than the QCD parameter $\Lambda$. The only case which
obeys this requirement is the interaction of two virtual
photons, $q^{2}=-Q^{2},\ \ p^{2}=-P^{2},\ \ Q,P>>\Lambda$. It is hardly
physical, but, for that, it allows to obtain a reliable information
about the high-energy behaviour of the cross-section. Later we shall
discuss uncertainties which arise as one goes  from large $Q$ and
$P$ down to values of the order of $\Lambda$.

According to (15) at $\nu\rightarrow\infty$ the dominant contribution
comes from the "ground state" in the integral (or/and sum) over $\kappa$
with the maximal trajectory $\alpha$. In the theory with a running
coupling constant this  may be a discrete level (for fixed total
momentum) or the beginning of a continuous spectrum, depending on the
chosen behaviour of the function $\eta$ in the infrared region. In the
first case  the Green function may
be approximated by \beq g^{4}f(\nu,q,q_{1},q'_{1})= \nu^{\alpha(q)-1}
\phi_{q}(q_{1},q_{2})\phi^{\ast}_{q}(q'_{1},q'_{2})
\eeq
The pomeron trajectory now has a nonzero slope
\beq
\alpha(q)=\alpha(0)-\alpha'q^{2},\ \ q\rightarrow 0
\eeq
The function $\phi_{q}(q_{1},q_{2})$ is the pomeron wave function in the
momentum space defined and normalized according to (13) and (14). From
(19) one concludes that small values of $q$, $q^{2}\sim 1/\alpha'\ln\nu$
 are essential at large $\nu$ so that one can safely take the wave
functions at $q=0$ in (19):
\beq
g^{4}f(\nu,q,q_{1},q'_{1})= \nu^{\alpha(q)-1}
\phi_{0}(q_{1},q_{2})\phi^{\ast}_{0}(q'_{1},q'_{2})
\eeq

With (20) and (21) the integration over $q$ becomes trivial in (5).
Doing the
$q_{1}$ and $q'_{1}$ integrals and changing the integration over $R^{2}$
for that over $z$ we get
\beq
A(q,p)=4i\nu e^{4}N_{c}^{2}2\alpha'\ln\nu (2\pi)^{-3}
\int_{0}^{\infty}rdr\rho_{q}(r)\int_{0}^{\infty}r'dr'\rho_{p}(r')
\int_{0}^{z_{0}(r,r')}(dz/z)(1-\exp (-z))
\eeq
where
\beq
z_{0}(r,r')=\frac{\nu^{\Delta}}{16\pi\alpha'\ln\nu}
(\phi_{0}(r)-\phi_{0}(0))(\phi_{0}(r')-\phi_{0}(0))^{\ast}
\eeq
and $\Delta=\alpha(0)-1$. If the ground state belongs to the continuous
spectrum, as in the BFKL theory, then $\alpha(0)$ depends on the
parameter $\kappa$: $\alpha_{\kappa}(0)\simeq\alpha(0)-a\kappa^{2}$.
Integration over $\kappa$ leads to (23) with an extra factor
$\sqrt{\pi/a\ln\nu}$. This will not have any significant influence for
the following, so that we assume (23) to be valid.

The dominant contribution to the integrals over $r$ and $r'$ comes from
the region of small $Qr$ and $Pr'$. The exact way the difference
$\phi_{0}(r)-\phi_{0}(0)$ goes to zero as $r\rightarrow 0$ has to be
studied
from the equation (8) for $\psi$ and subsequent transition to $\phi$
via Eq. (13) and then to the transverse space. It is however not very
important for the following. We assume
\beq
\phi_{0}(r)-\phi_{0}(0)=ar^{\beta},\ \ r\rightarrow 0
\eeq
Since ${\rm dim}\phi(r)=-1$, in the BFKL theory with a fixed coupling,
one naturally finds $\beta=1$. For a running coupling the equation is
better behaved at small $r$, so that we expect $\beta\geq1$ (there also
may appear logarithms in (24))
With (24) at small $r$ and $r'$
\beq
z_{0}(r,r')=\frac{a^{2}\nu^{\Delta}}{16\pi\alpha'\ln\nu}
(rr')^{\beta}
\eeq

Evidently the asymptotic behaviour of the amplitude (22) depends on the
magnitude of the parameter
\beq
\xi=\frac{a^{2}\nu^{\Delta}}{16\pi\alpha'\ln\nu (QP)^{\beta}}
\eeq
If $\xi\rightarrow 0$ then the exponential in (22) can be developed in a
power series and only the first term contributes. This means that the
amplitude and the cross-section reduce to a single pomeron exchange.
This case is that of a relatively large $x=q^{2}/2\nu$ and
$y=P^{2}/2\nu$
\beq
\nu^{2\Delta/\beta-2}<<xy<<1
\eeq
(recall that $\Delta$ is asssumed small in the theory, so that the two
inequalities in (27) are compatible).
Passing in (22) to integrations over $Qr$ and $Pr'$ we find in this case
\beq
A(q,p)=i\nu e^{4}N_{c}^{2}a^{2}B^{2} (2\pi)^{-4}
\frac{\nu^{\Delta}}{(QP)^{\beta}}
\eeq
where $B$ is result of the $r$ or $r'$ integration
\beq
B= \int_{0}^{\infty}r^{1+\beta}dr\rho_{q_{0}}(r),\ \ q_{0}^{2}=1
\eeq
The total cross-section is obtained from (28) dividing it by $4i\nu$.
It rises with the c.m energy squared $s=2\nu$ as a power $s^{\Delta}$

In the opposite case $\xi\rightarrow\infty$ all multipomeron exchanges
contribute. In the integral (22) regions of small $Qr$ and $Pr'$ give
the dominant contribution. So using (18) we find for the leading term
\[
A(q,p)=4iC\nu\alpha'\ln\nu\]\beq
\int_{0}^{1/2}dr/(r\ln(1/r^{2}))\int_{0}^{1/2}dr'/(r'\ln(1/{r'}^{2}))
\int_{0}^{1}(dz/z)(1-\exp (-\xi z(rr')^{\beta}))
\eeq
The constant factor $C$ is given by
\beq
C=(4/9)e^{4}N_{c}^{2}(2\pi)^{-3} (\sum_{f}Z_{f}^{2})^{2}
\eeq
The asymptotical behaviour of (30) at $\xi>>1$ is easily found to be
\beq
A(q,p)\simeq i\beta^{2}C\alpha'\nu\ln\nu\ln\xi(\ln\ln\xi)^{2}
\eeq
As a function of energy for  fixed $q$ and $p$ the corresponding
 cross-section  behaves as presented in Eq. (1).

\section{Real particles}

Now we pass to the more realistic case when one (the target) or both of
the interacting particles are physical, not virtual. For the
photon-photon interaction considered before it means that either
$p^{2}=0$ or both $q^{2}=p^{2}=0$. In the first case we are dealing with
the structure function, which is related to the $\gamma^{\ast}$-target
cross-section by means of the relation
\beq
F_{2}(x,Q^{2})=Q^{2}(\sigma_{T}+\sigma_{L})/\pi e^{2}
\eeq
Here $\sigma_{T}$ is the cross-section for the transversely polarized
projectile photon , studied in Sec. 3, and $\sigma_{L}$ is the analogous
cross-section for the longitudinally polarized projectile photon, which
differs from $\sigma_{T}$ by the substitution of the longitudinal photon
colour density
\beq
\rho_{q,L}(r)=4Q^{2}
\sum_{f=1}^{N_{f}}Z_{f}^{2}\int_{0}^{1}d\alpha
\alpha^{2}(1-\alpha)^{2}
{\rm K}_{0}^{2}(\epsilon_{f} r)
\eeq
instead of the transversal one (3).

With real particles the momenta associated with their structure become
comparable to $\Lambda$ and non-perturbative effects appear. For the
structure function they are limited to the target part. With the photon
as a target, in particular, we have to put $p^{2}=0$, so that the
momenta in the target quark loop become small and we can no more use
 the perturbative treatment. Not only the expression (3) for the colour
density becomes invalid, but also the factorizability of the interaction
with gluons becomes questionable for the target. The natural scale for
the intergluon distance $r'$ in the target becomes $1/m$ where evidently
$m=\Lambda$. The characteristic parameter $\xi$ then is, instead of (26),
\beq
\xi=\frac{a^{2}\nu^{\Delta}}{16\pi\alpha'\ln\nu (mQ)^{\beta}}
\eeq

We again observe two different regimes. If $\xi$ is small then the
single pomeron exchange dominates. In this case  values
of $r'$ of the order $1/\Lambda$ are essential, where we do not know the
explicit form of the photon colour density. It however enters in
the amplitude only  integrated over all $r'$,
through the number $B$ given by (29). Therefore, although the numerical
value of $B$ will be changed by non-perturbative effects at $p^{2}=0$,
Eq. (28) for the amplitude will be left intact, with the substitution
$P\rightarrow m=\Lambda$. Thus in the region
\beq
(Q/\Lambda)^{2-\Delta/\beta}<<x<<1
\eeq
the structure function is described by the single pomeron exhange and
behaves as
\beq
 F_{2}(x,Q^{2})\sim\nu^{\Delta}Q^{2-\beta}\sim x^{-\Delta}
Q^{2+2\Delta-\beta}
\eeq

If $x$ gets still smaller and $\xi\rightarrow\infty$ then we have to sum
over all pomeron exchanges. From the structure of the eikonal function
we observe that now small values of $r'<<1/\Lambda$ contribute. For such
values of $r'$ we expect that the target colour density will still be
given by its perturbative expression (3), since the coupling constant
becomes small. Then the amplitude  retains its eikonal form.  At
small distance $r'$ the behaviour of
$\rho_{p}(r')$ will be given by the same Eq. (18) in which we only have
to substitute $q$ by the scale $m$, which is now
the smallest quark mass. As a result in the region
\beq
x<<(Q/m)^{2-\Delta/\beta}
\eeq
the asymptotical behaviour of the amplitude has the same form (32) as
with a virtual photon target, with $\xi$ now given by (35).
Correspondingly the photon structure function has the behaviour
\beq
F_{2}^{(\gamma)}(x,Q^{2})\sim Q^{2}\ln\nu \ln\xi (\ln\ln\xi)^{2}
\eeq
It strongly violates scaling: for fixed $x$ it rises roughly as $Q^{2}$.
As a function of $x$ it rises as
$(\ln (1/x)\ln\ln (1/x))^{2}$.

One may ask what will happen if instead of the photon target we take a
hadronic one. In the region (36), where the single pomeron exchange
dominates, we do not find any change, except that the corresponding
hadronic colour density should be integrated in (29) for the target
factor $B$. In the small $x$ region (38) the answer is more speculative.
We expect that for a hadronic target the coupling to the external
particle is softer than for the elementary photon, so that the colour
density $\rho_{p}(r')$ becomes less singular as $r'\rightarrow 0$. Then
finite values of $r'\sim 1/\Lambda$ will give the dominant contribution
also to the multiple pomeron exchange. We do not know the structure of
the hadronic coupling to gluons in this region. The factorizabilty of
this coupling and therefore the eikonal structure of the amplitude
become  questionable. If we assume that the factorizability still
persisits and the amplitude continues to have an eikonal form (2) then
we can easily find its asymptotic behaviour. It is given by Eq. (32) in
which a factor $(1/2)\beta\ln\xi$ should be substituted by the integral
of the hadronic colour density (assumed to exist)
\beq
B_{0}=\int_{0}^{\infty}rdr\rho_{p}(r)
\eeq
Thus we obtain for the $\gamma^{\ast}-h$ amplitude in the region (38)
\beq
A(q,p)\simeq 2i\beta B_{0}C\alpha'\nu\ln\nu\ln\xi\ln\ln\xi
\eeq
It rises with $Q$ roughly as $Q^{2}$ and with $x$ as
$(\ln (1/x))^{2}\ln\ln (1/x)$.

Finally we briefly discuss  cross-sections for the interaction of real
particles, both the target and projectile. With the appropriate scale
$m$ in the target and projectile, the characteristic parameter $\xi$ is
now given by \beq
\xi=\frac{a^{2}\nu^{\Delta}}{16\pi\alpha'\ln\nu m^{2\beta}}
\eeq
and is always large. Therfore all pomeron exchanges contribute. A
rigorous result then may be obtained only for the real photon target
and projectile. In this case small values of $r$ and $r'$ give the
dominant contribution, where the colour densities continue to be given
by their perturbative expressions. As a result the amplitude retains its
eikonal form. Its asymptotcal behaviour will be given by the same
Eq. (32) with $\xi$ given by (42) and the scale $m$ taken as the minimal
quark mass. The resulting cross-section rises as indicated in Eq. (1)
and weakly violates the Froissart bound.

Changing one or both interacting particles to hadrons will bring in
values of $r$ or/and $r'$ of the order $1/\Lambda$ and
consequently non-perturbative effects. Then the factorizability of the
gluon coupling to the target or/and projectile and eikonalization become
a problem. If one assumes that they still persist, then the asymptotical
behaviour of the amplitude and the cross-section will differ from the
pure photonic ones by the substitution of the factor
$(1/2)\beta\ln\ln\xi$  by the integral of the density $B_{0}$ (Eq. (40))
for the target and/or the projectile. One has also to take into account
that the scale appropriate for a hadron is $m=\Lambda$. As a result the
purely hadronic cross-section will rise as $(\ln s)^{2}$ in accordance
with the Froissart bound and in the same manner as in the old Froissaron
model of [ 3 ].

\section{Conclusions}

We have studied the high-energy behaviour of the cross-sections for the
scattering of weakly virtual and real particles in the framework of the
theory of reggeized gluons, with $N_{c}\rightarrow\infty$ and a running
coupling constant. The results are more or less rigorous for interacting
photons and more speculative for hadronic targets and projectiles.

As $x$ gets smaller we observe a transition from a single pomeron
exchange dominance to multiple pomeron exchanges, which sum into an
eikonal amplitude. In both regimes we find strong violation of scaling:
the structure function rises as $Q$ and $Q^{2}$ for the two regimes
respectively. As a function of $x$ it changes its behaviour from
a power growth $1/x^{\Delta}$ to a logarithmic one
$(\ln (1/x))^{2}(\ln\ln(1/x))^{k}$, $k=1$ or $2$.

The cross-section for the scattering of real photons grows as
$(\ln s\ln\ln s)^{2}$ and violates the Froissart bound. For hadronic
projectiles and targets this violation is absent, in all probability.
However eikonalization of the hadronic amplitude cannot be demonstrated
because of non-perturbative effects.

Most of the results are based only on the existence of a nonzero pomeron
slope for a running coupling. The details of the introduction of the
running coupling seem unimportant. However the elimination of gluonic
configurations different from multipomeron ones rests heavily on the
bootstrap condition, which must be preserved. The only other property of
the pomeron which enters the asymptotical behaviour of the amplitudes is
the behaviour of its wave function at small distances $r$. In particular
it determines the value of $x$ at which multiple pomeron exchanges begin
to give a sizable contribution. The small $r$ behaviour of the pomeron
wave function has to be studied from the corresponding equation with the
running coupling constant. This problem is currently under study.

\section {Acknowledgements}

The author is grateful to Lev Lipatov and Misha Ryskin for valuable
discussions.

 \newpage

\section{References}

1. M.A.Braun, preprint SPbU-IP-1995/3, HEP-PH/9502403.\\
2. V.S.Fadin, E.A.Kuraev and L.N.Lipatov, Phys. Lett. {\bf B60} (1975)
50.\\
I.I.Balitsky and L.N.Lipatov, Sov.J.Nucl.Phys. {\bf 15} (1978) 438.\\
3.A.Capella, J.Kaplan and J.Tran T.V., Nucl. Phys. {\bf B10} (1969) 519;\\
B.Z.Kopeliovich and L.I.Lapidus, Sov.Phys. JETP {\bf 43} (1976) 32;\\
M.S.Dubovikov and K.A.Ter-Martirosyan, Nucl. Phys. {\bf B124} (1977)
 163;\\
M.S.Dubovikov et al., {\it ibid} {B123 (1977) 143.\\
4. L.N.Lipatov, Yad. Fiz. {\bf 23} (1976) 642.\\
5. M.A.Braun, Phys.Lett. {\bf B 345} (1995) 155; preprint US-FT/11-94\\

 \end{document}

